\documentclass[10pt,letterpaper]{article}
\usepackage[top=0.85in,left=2.75in,footskip=0.75in,marginparwidth=2in]{geometry}
\usepackage[utf8]{inputenc}
\usepackage{lmodern}
\usepackage{amsmath}
\usepackage{cite}

\usepackage{nameref,hyperref}

\usepackage[right]{lineno}

\usepackage{microtype}
\DisableLigatures[f]{encoding = *, family = * }

\raggedright
\usepackage{changepage}

\usepackage[aboveskip=1pt,labelfont=bf,labelsep=period,singlelinecheck=off]{caption}

\makeatletter
\renewcommand{\@biblabel}[1]{\quad#1.}
\makeatother

\usepackage{lastpage,fancyhdr,graphicx}
\usepackage{epstopdf}
\fancyheadoffset[L]{2.25in}
\fancyfootoffset[L]{2.25in}

\usepackage{color}

\definecolor{Gray}{gray}{.25}

\usepackage{graphicx}

\usepackage{sidecap}

\usepackage{wrapfig}
\usepackage[pscoord]{eso-pic}
\usepackage[fulladjust]{marginnote}
\reversemarginpar
\usepackage{geometry}
\begin{document}
\vspace*{0.35in}

\begin{flushleft}
{\Large
\textbf\newline{BowTie - A deep learning feedforward neural network for sentiment analysis}
}
\newline
\\
Apostol Vassilev
\\
\bigskip
{\bf National Institute of Standards and Technology}

\bigskip
apostol.vassilev@nist.gov
\end{flushleft}
\begin{abstract}
How to model and encode the semantics of human-written text and select the type of neural network to process it are not settled issues in sentiment analysis. Accuracy and transferability are  critical issues in machine learning in general. These properties are closely related to the loss estimates for the trained model. I present a computationally-efficient and accurate feedforward neural network for sentiment prediction capable of maintaining low losses. When coupled with an effective semantics model of the text, it provides highly accurate models with low losses. Experimental results on representative benchmark datasets and comparisons to other methods\footnote{DISCLAIMER: This paper is not subject to copyright in the United States. Commercial products are identified in order to adequately specify certain procedures. In no case does such identification imply recommendation or endorsement by the National Institute of Standards and Technology, nor does it imply that the identified products are necessarily the best available for the purpose.}show the advantages of the new approach. 
\end{abstract}

\section*{Introduction}
When approaching the problem of applying deep learning to sentiment analysis one faces at least five classes of issues to resolve. First, what is the best way to encode the semantics in natural language text so that the resulting digital representation captures well the semantics in their entirety and in a way that can be processed reliably and efficiently by a neural network and result in a highly accurate model? This is a critically important question in machine learning because it directly impacts the viability of the chosen approach. There are multiple ways to encode sentences or text using neural networks, ranging from a simple encoding based on treating words as atomic units represented by their rank in a vocabulary~\cite{Brants-at-al-2007}, to using word embeddings or distributed representation of words~\cite{Mikolov-at-al-2013}, to using sentence embeddings. Each of these encoding types have different complexity and rate of success when applied to a variety of tasks. The simple encoding method offers simplicity and robustness.  The usefulness of word embeddings has been established in several application domains, but it is still an open question how much better it is than simple encoding in capturing the entire semantics of the text in natural language processing (NLP) to provide higher prediction accuracy in sentiment analysis. Although intuitively one may think that because word embeddings do capture some of the semantics contained in the text this should help, the available empirical test evidence is inconclusive. Attempts to utilize sentence embeddings have been even less successful~\cite{Conneau-at-al-2017}.         

Second, given an encoding, what kind of neural network should be used? Some specific areas of applications of machine learning have an established leading network type. For example, convolutional neural networks are preferred in computer vision. However, because of the several different types of word and sentence encoding in natural language processing (NLP), there are multiple choices for neural network architectures, ranging from feedforward  to convolutional and recurrent neural networks.   

Third, what dataset should be used for training? In all cases the size of the training dataset is very important for the quality of training but the way the dataset is constructed and the amount of meta-data it includes also play a role. For example, the Keras IMDB Movie reviews Dataset~\cite{Keras:IMDB} (KID) for sentiment classification contains human-written movie reviews. A larger dataset of similar type is the Stanford Large Movie Review Dataset (SLMRD)~\cite{Stanford:LMRD}. I consider KID and SLMRD in detail in Sections~\ref{sec:KID} and \ref{sec:SLMRD}. Generally, simpler encodings and models trained on large amounts of data tend to outperform complex systems trained on smaller datasets~\cite{Mikolov-at-al-2013}.

Fourth, what kind of training procedure should be employed - supervised or unsupervised? Traditionally, NLP systems are trained on large unsupervised corpora and then applied on new data. However, researchers have been able to leverage the advantages of supervised learning and transfer trained models to new data by retaining the transfer accuracy~\cite{Conneau-at-al-2017}.

Fifth, when training a model for transfer to other datasets, what are the model characterizing features that guarantee maintaining high/comparable transfer accuracy on the new dataset? Certainly, training and validation accuracy are important but so are the training and validation losses. Some researchers argue that the gradient descent method has an implicit bias that is not yet fully understood, especially in cases where there are multiple solutions that properly classify a given dataset~\cite{Nacson-at-al-2018}. Thus, it is important to have a neural network with low loss estimates for a trained model to hope for a good and reliable transfer accuracy.

The primary goal of this paper is to shed light on how to address these issues in practice. To do this, I introduce a new feedforward neural network for sentiment analysis and draw on the experiences from using it with two different types of word encoding: a simple one based on the word ranking in the dataset vocabulary; the other judiciously enhanced with meta-data related to word polarity. The main contribution of this paper is the design of the BowTie neural network in Section~\ref{sec:BowTie}.  

\section{Data encoding and datasets}
\label{sec:data}
As discussed above, there are many different types of encodings of text with different complexity and degree of effectiveness. Since there is no convincing positive correlation established in the literature between complexity of the encoding  and higher prediction accuracy, it is important to investigate the extent to which simple data encodings can be used for sentiment analysis. Simpler encodings have been shown to be robust and efficient~\cite{Brants-at-al-2007}. But can they provide high prediction accuracy in sentiment analysis? 

I investigate this open question by evaluating the accuracy one may attain using two types of text encoding on representative benchmark datasets. 
\subsection{Multi-hot encoding}
\label{ec:multihot}
\begin{wrapfigure}[16]{l}{70mm}
\includegraphics[width=3.1in]{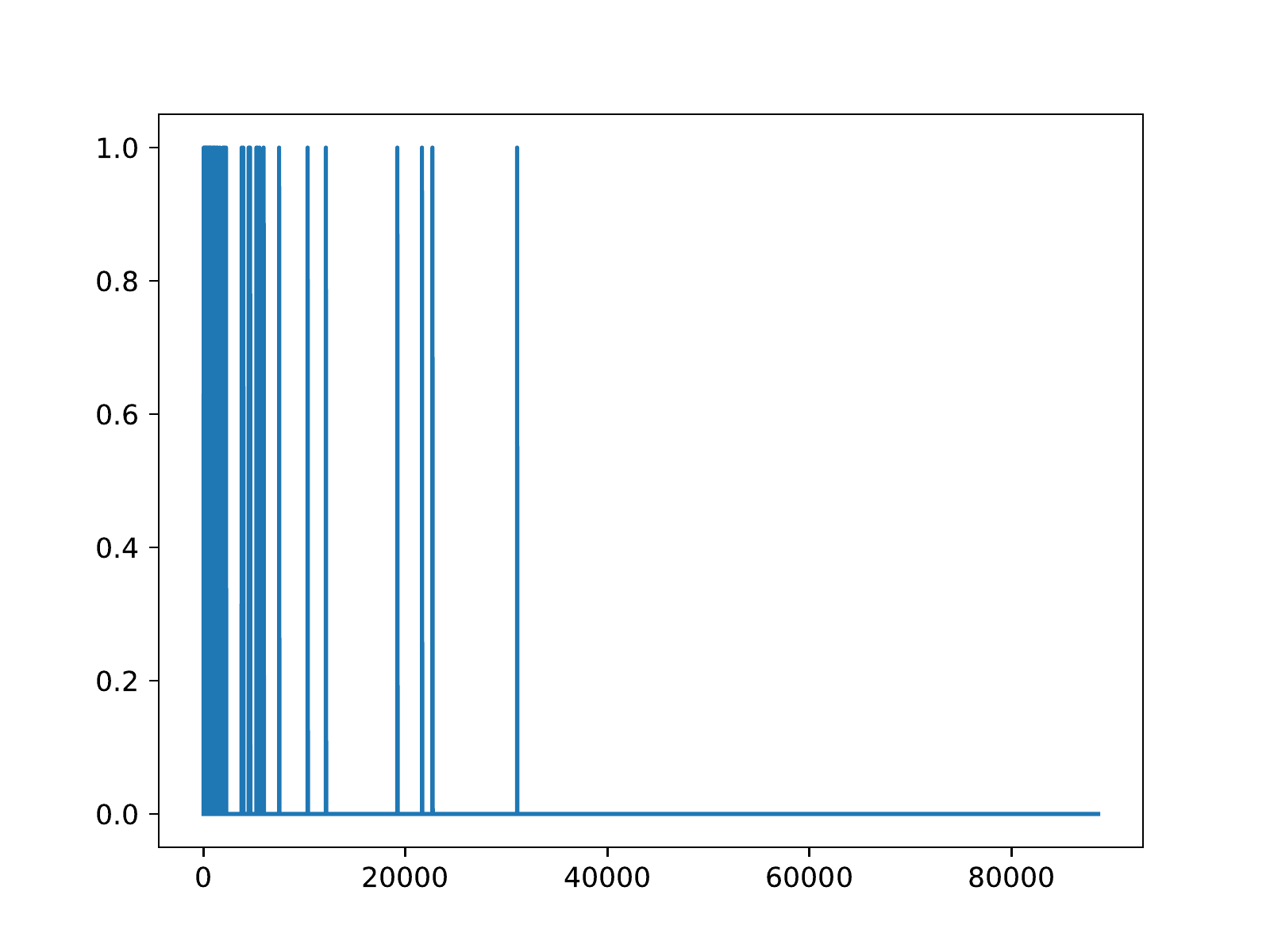}
\captionsetup{labelformat=empty} 
\caption{} 
\label{fig:multihot} 
\end{wrapfigure}

The first encoding is the {\em multi-hot} encoding of text~\cite{TensorFlow:Lib} which can be defined as follows.
\newpage
\marginpar{
\vspace{1.3cm} 
\color{Gray} 
\textbf{Figure \ref{fig:multihot}. A multi-hot encoded text in $D,\Pi_D$ with $M=88\,587$.} 
}
Let $\pi$ be a linguistic type (e.g. morpheme, word) and let $\Pi_D$ be the set of all such linguistic types in a dataset $D$. Let $M=\vert{\Pi_D}\vert$ be the cardinality of $\Pi_D$. Let $\psi$ be a linguistic text type (e.g., a movie review) and let $\Psi_D$ be the set of all texts in $D$. Let $N=\vert\Psi_D\vert$ be the cardinality of $\Psi_D$. Let $\Pi_D$ and $\Psi_D$ be finite sets such that the elements in each set are enumerated by $\{0,..., M\}$ and $\{0,...,N\}$ respectively. Let $T^{NxM}$ be a tensor of real numbers of dimensions $N$ by $M$, whose elements are set as follows:

\begin{align}
\label{eq:multihot}
  \left\{\tau_{jk}\right\} &=
  \begin{cases}
    \, 1, &\text{if $\pi_k \in \psi_j$;} \\[12pt]
    \, 0, &\text{otherwise.}
  \end{cases}
\end{align}

The multi-hot encoding (\ref{eq:multihot}) represents a very simple model of the semantics in $\psi$, $\forall \psi\in\Psi_D$. An example of a multi-hot encoded text is shown in Figure~\ref{fig:multihot}.

\subsection{Polarity-weighted multi-hot encoding}
\label{sec:wmultihot}
The second encoding I consider is similar to the multi-hot encoding in the sense it has the same non-zero elements but their values are weighted by the cumulative effect of the polarity of each word present in a given text $\pi$, as computed by~\cite{Potts-2011}. Let $c_{\pi, \psi}$ be the number of tokens of the linguistic type $\pi$ in a text $\psi$. Let $\xi_\pi$ be the polarity rating of the token $\pi\in \Pi_D$. Naturally, I assume that if $\Xi_D$ is the set of all polarity ratings for tokens $\pi\in\Pi_D$, then $\vert\Xi_D\vert = \vert\Pi_D\vert$. Let $\omega_{\xi\pi\psi} = \xi_\pi*c_{\pi, \psi}$ be the cumulative polarity of $\pi$ in the text $\psi$. Let $\Omega_D=\{\omega_i\}_{i=0}^M$ and $C_D=\{c_i\}_{i=0}^M$.   

\begin{wrapfigure}[14]{l}{70mm}
\includegraphics[width=3.1in]{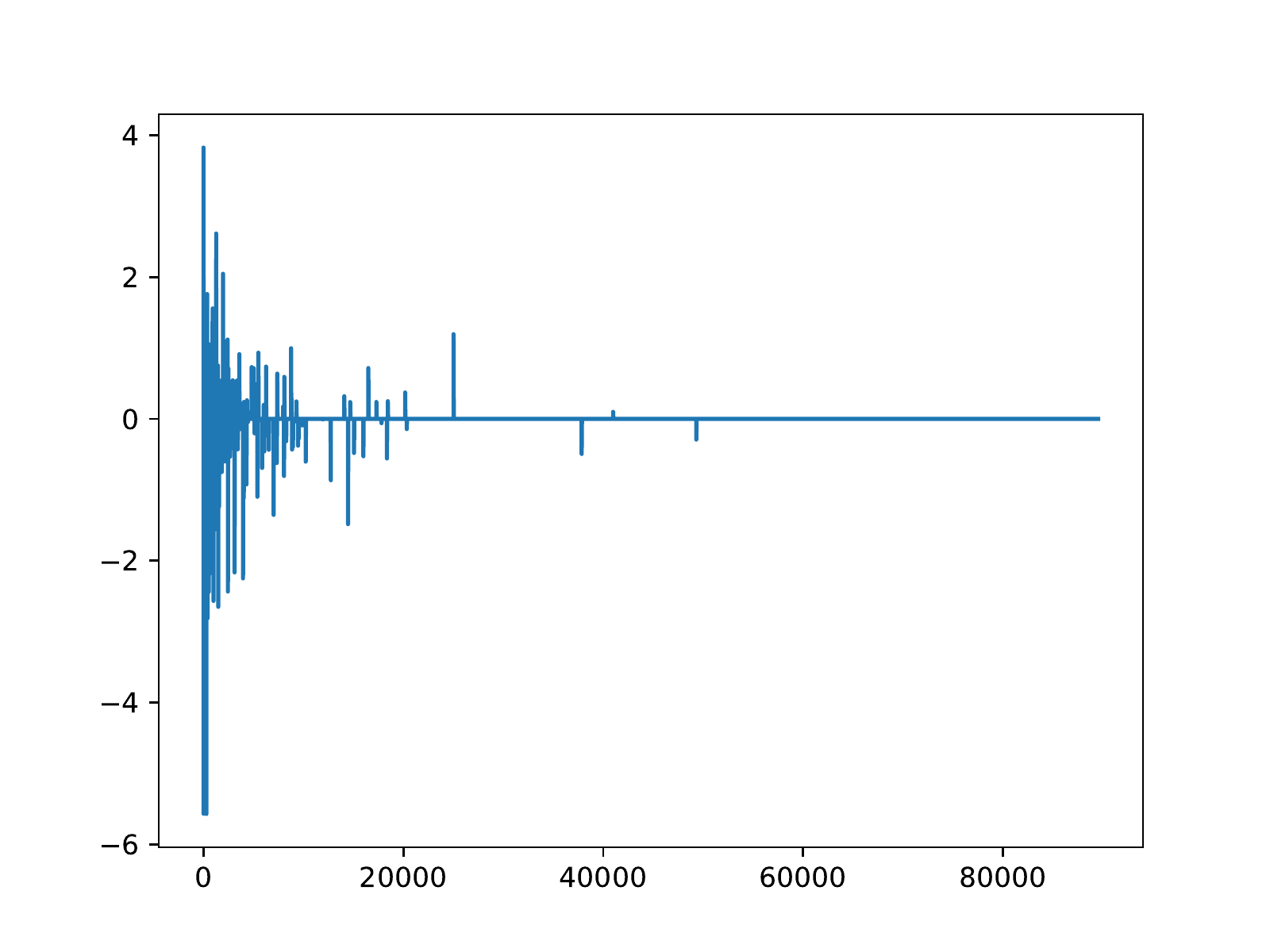}
\captionsetup{labelformat=empty} 
\caption{} 
\label{fig:wmultihot} 
\end{wrapfigure}

\marginpar{
\vspace{1.3cm} 
\color{Gray} 
\textbf{Figure \ref{fig:wmultihot}. A polarity-weighted multi-hot encoded text in $D, \Pi_D$ with $M=89\, 527$.} 
}
Let $\Theta^{NxM}$ be a tensor of real numbers of dimensions $N$ by $M$, whose elements are set as follows:
\begin{align}
\label{eq:polarity-weighted-multihot}
  \left\{\theta_{jk}\right\} &=
  \begin{cases}
    \, \omega_{\xi_k\pi_k\psi_j}, &\text{if $\pi_k \in \psi_j$;} \\[12pt]
    \, 0, &\text{otherwise.}
  \end{cases}
\end{align}
The polarity-weighted multi-hot encoding (\ref{eq:polarity-weighted-multihot}) represents a more comprehensive  model of the semantics in $\psi$, $\forall \psi\in\Psi_D$, that captures more information about $\psi$. I will attempt to investigate if and how much this additional information helps to improve the sentiment predictions in Section~\ref{sec:results}.  An example of a polarity-weighted multi-hot encoded text is shown in Figure~\ref{fig:wmultihot}.

\subsection{The Keras IMDB Dataset (KID)}
\label{sec:KID}
The KID~\cite{Keras:IMDB} contains 50 000 human-written movie reviews that are split in equal subsets of 25 000 for training and testing and further into equal categories labeled as positive or negative. For convenience, the reviews have been pre-processed and each review is encoded as a sequence of integers, representing the ranking of the corresponding word in $\Pi_D$ with $\vert\Pi_D\vert=88\,587$. As such, it can be easily encoded by the multi-hot encoding (\ref{eq:multihot}). 

\subsection{The Stanford Large Movie Review Dataset (SLMRD)}
\label{sec:SLMRD}
SLMRD contains 50 000 movie reviews, 25 000 of them for training and the rest for testing. The dataset comes also with a processed bag of words and a word polarity index~\cite{Stanford:LMRD, maas-EtAl:2011:ACL-HLT2011}. SLMRD contains also 50 000 unlabeled reviews intended for unsupervised learning. It comes with a $\Pi_D$, polarity ratings $\Omega_D$, and word counts $C_D$ with $\vert\Omega_D\vert=\vert C_D\vert=\vert\Pi_D\vert = 89\, 527$. 

\begin{wrapfigure}[15]{l}{60mm}
\includegraphics[width=60mm]{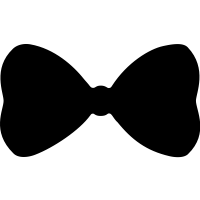}
\captionsetup{labelformat=empty} 
\caption{} 
\label{fig:theBowTie} 
\end{wrapfigure}

\section{The BowTie\protect\footnote{The name is inspired by the classic image of the bow tie - see Figure~\ref{fig:theBowTie}} feedforward neural network}
\label{sec:BowTie}
\marginpar{
\vspace{1.3cm} 
\color{Gray} 
\textbf{Figure \ref{fig:theBowTie}. The classic bow tie.} The bow tie originated among Croatian mercenaries during the Thirty Years' War of the 17th century. It was soon adopted by the upper classes in France, then a leader in fashion, and flourished in the 18th and 19th centuries.  (Wikipedia)
}
As discussed above, the ability of the network to provide accurate predictions and maintain low losses is very important for allowing it to transfer to other datasets and maintain the same or higher prediction accuracy as on the training dataset. 

I now introduce a feedforward neural network with that criteria in mind. By way of background~\cite{Goodfellow-et-al-2016}, logistic regression computes the probability of a binary output $\hat{y}_i$ given an input $x_i$ as follows:

\begin{equation}
P(\hat{\textbf{y}}\vert \textbf{X}, \textbf{w}) = \prod_{i=1}^n \textbf{Ber}[\hat{y}_i\vert \textbf{sigm}(x_i\textbf{w})],
\end{equation}

where $\textbf{Ber[ ]}$ is the Bernouli distribution, $\textbf{sigm}()$ is the sigmoid function, $\textbf{w}$ is a vector of weights. The cost function to minimize is $\textbf{C}(\textbf{w})=-\log P(\hat{\textbf{y}}\vert \textbf{X}, \textbf{w})$. This method is particularly suitable for sentiment prediction. One critical observation is that logistic regression can be seen as a special case of the generalized linear model. Hence, it is analogous to linear regression. In matrix form, linear regression can be written as

\begin{equation}
\hat{\textbf{y}} = \textbf{X}\textbf{w} + \boldsymbol{\epsilon},
\end{equation}    

where $\hat{\textbf{y}}$ is a vector of predicted values $\hat{y}_i$ that the model predicts for $\textbf{y}$, $\textbf{X}$ is a matrix of row vectors $x_i$ called regressors, \textbf{w} are the regression weights, and $\boldsymbol{\epsilon}$ is an error that captures all factors that may influence $\hat{\textbf{y}}$ other than the regressors $\textbf{X}$.

The gradient descent algorithm used for solving such problems~\cite{Goodfellow-et-al-2016} may be written as
\begin{equation}
\label{eq:gd}
\textbf{w}^{(k+1)} = \textbf{w}^{(k)} - \rho^{(k)}\textbf{g}^{(k)} +\boldsymbol{\epsilon}^{(k)},
\end{equation}
where $\textbf{g}^{(k)}$ is the gradient of the cost function $\textbf{C}(\textbf{w})$, $\rho^{(k)}$ is the learning rate or step size, and $\boldsymbol{\epsilon}^{(k)}$ is the error at step $k$ of the iterative process. 
 One error-introducing factor in particular is the numerical model itself and the errors generated and propagated by the gradient descent iterations with poorly conditioned matrices run on a computer with limited-precision floating-point numbers. Even if regularization is used, the specific parameters used to weigh them in the equation (e.g., the $L_2$-term weight or the dropout rate) may not be optimal in practice thus leading to potentially higher numerical error. This is why it is important to look for numerical techniques that can reduce the numerical error effectively. This observation inspires searching for techniques similar to multigrid from numerical analysis that are very effective at reducing the numerical error~\cite{Bramble-1993}.

\subsection*{The neural network design}

\begin{figure}[ht] 


\includegraphics[width=\textwidth]{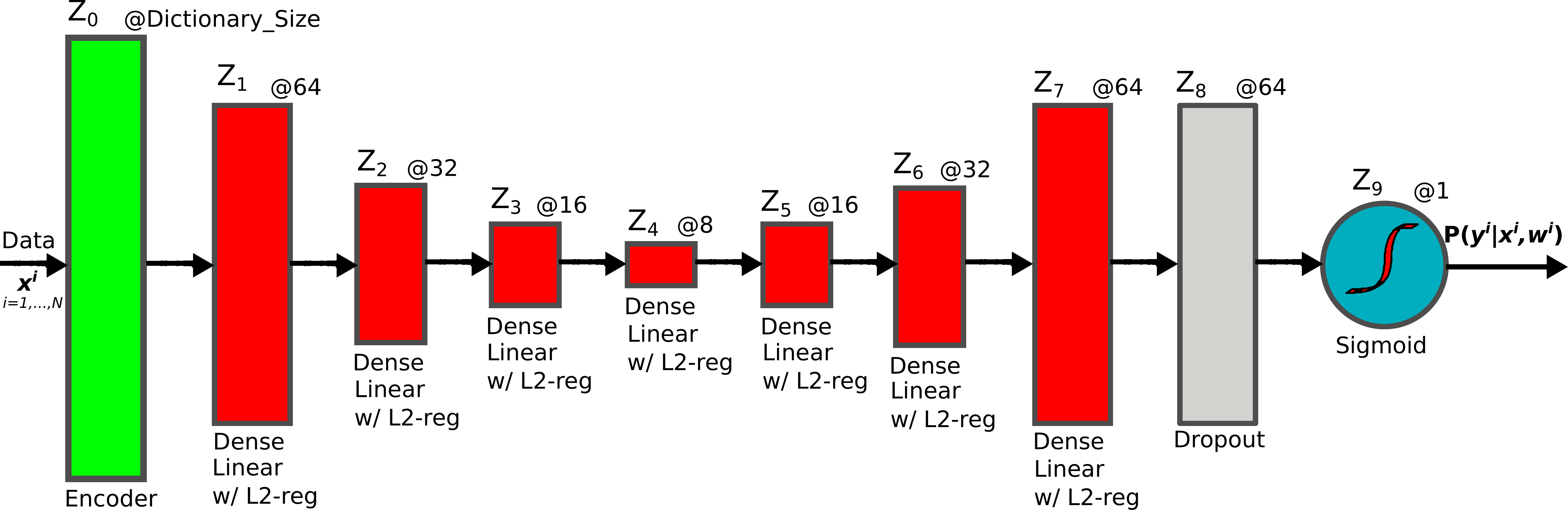}

\caption{\color{Gray} \textbf{The BowTie neural network.} The estimated probability $P(\hat{\textbf{y}}^{(i)}|\textbf{x}^{(i)}, \textbf{w}^{(i)})$ may be fed into a post-processing discriminator component to assign a category (pos/neg) for the input $\textbf{x}^{(i)}$ with respect to a discriminator value $\delta\in[0, 1]$. All experiments presented in this paper use $\delta=0.5$.
} 
\label{fig:BowTie} 

\end{figure}
The feedforward neural network in  Figure \ref{fig:BowTie} consists of one encoding layer, a cascade of dense linear layers with $L_2$-regularizers and of appropriate output size followed by a dropout regularizer and a sigmoid. The encoder takes care of encoding the input data for processing by the neural network. In this paper I experiment with the two encodings defined in Section~\ref{sec:data}: the simple multi-hot encoding and the polarity-weighted multi-hot encoding. 

The sigmoid produces the estimated output probability $P(\hat{\textbf{y}}^{(i)}|\textbf{x}^{(i)}, \textbf{w}^{(i)})$, which may be used to compute the negative log-loss or binary cross-entropy as 
\begin{equation}
-\left[\textbf{y}\log(P(\hat{\textbf{y}}^{(i)}|\textbf{x}^{(i)}, \textbf{w}^{(i)}))+(1-\textbf{y})\log(1-P(\hat{\textbf{y}}^{(i)}|\textbf{x}^{(i)}, \textbf{w}^{(i)}))\right]. 
\end{equation}

The binary cross-entropy provides a measure for quality and robustness of the computed model. If the model predicts correct results with higher probability, then the binary-cross entropy tends to be lower. If however the model predicts correct results with probability close to the discriminator value or predicts an incorrect category, the binary cross-entropy tends to be high. Naturally, it is desirable to have models that confidently predict correct results. It is also important to have models that maintain low binary cross-entropy for many training epochs because, depending on the training dataset, the iterative process (\ref{eq:gd}) may need several steps to reach a desired validation accuracy. Models that quickly accumulate high cross-entropy estimates tend to overfit the training data and do poorly on the validation data and on new datasets.   

\paragraph{Hyperparameters.} There are several hyperparameters that influence the behavior of BowTie, see Table~\ref{tab:Hyperparameters}.
\begin{table}[!ht]
\begin{adjustwidth}{0in}{0in} 
\centering
\caption{{\bf BowTie hyperparameters.}}
\begin{tabular}{|c|c|}
\hline
\multicolumn{2}{|c|}{\bf Hyperparameters} \\ \hline
{\bf name}& {\bf values/range}\\ \hline
$L_2$-regularization weight & 0.01 - 0.02\\ \hline
Dropout rate & 0.2 - 0.5\\ \hline
Optimizer & NADAM, ADAM, or RMSProp\\ \hline
Dense Layer Activation & None (Linear network), RELU\\ \hline
\end{tabular}
\label{tab:Hyperparameters}
\end{adjustwidth}
\end{table}
For optimal performance the choice of the dense layer activation should be coordinated with the choice for the $L_2$-regularization weight. This recommendation is based on the computational experience with BowTie and is in line with the findings in \cite{Hayou-at-al-2019} about the impact of the activation layer on the training of neural networks in general. The linear network (no dense layer activation) runs better with $L_2$-regularization weight close to 0.02 but rectified linear unit (RELU) activation runs better with $L_2$-regularization weight close to 0.01. The network can tolerate a range of dropout rates but a dropout rate of 0.2 is commonly recommended in the literature and works well here too. The choice of the optimizer can affect the learning rate, the highest accuracy attained and the stability over several epochs. BowTie performs well with adaptive momentum (ADAM), Nesterov adaptive momentum (NADAM) and Root Mean Square Propagation (RMSPRop), no dense layer activation, $L_2$-regularization set to 0.019 (this value resulting from hyperparameter optimization) and a dropout rate of 0.2. It is interesting to note that RMSProp tends to converge faster to a solution and sometimes with a higher validation accuracy than NADAM but the transfer accuracy of the models computed with NADAM tends to be higher than for models computed with RMSPRop. For example, a model trained on SLMRD with validation accuracy of 89.24 \%, higher than any of the data in Table~\ref{tab:Sec3-4} below, yielded 91.04 \% transfer accuracy over KID, which is lower than the results in Table~\ref{tab:Sec3-4}. This experimental finding is consistent over many tests with the two optimizers and needs further investigation in future research to explore the theoretical basis for it. 

\section{Training and transfer scenarios}
\label{sec:scenarios}
This section defines the objectives for the testing of the BowTie neural network shown in Figure~\ref{fig:BowTie} in terms of four training and transfer scenarios. 

But first it is important to decide on the type of training to employ - supervised or unsupervised. I embark on supervised training based on the findings in~\cite{Conneau-at-al-2017} about the advantages of supervised training and the availability of corpora of large labeled benchmark datasets~\cite{Keras:IMDB} and \cite{Stanford:LMRD}. 

These are the scenarios to explore:
\begin{itemize}
\item \textbf{Scenario 1 (Train and validate):} Explore the accuracy and robustness of the BowTie neural network with the simple multi-hot encoding by training and validating on KID.
\item \textbf{Scenario 2 (Train and validate):} Explore the accuracy and robustness of the BowTie neural network with the simple multi-hot encoding  by training and validating on SLMRD.
\item \textbf{Scenario 3 (Train and validate):} Explore the accuracy and robustness of the BowTie neural network with the polarity-weighted multi-hot encoding  by training and validating on SLMRD.
\item \textbf{Scenario 4 (Train, validate, and transfer):} Explore the transfer accuracy of the BowTie neural network with polarity-weighted multi-hot encoding by training on SLMRD and predicting on KID.
\end{itemize} 

The primary goal of this exploration is to establish some baseline ratings of the properties of the BowTie neural network with the different encodings and compare against similar results for other neural networks with other types of encoding. This provides a quantitative criteria for comparative judging.  
\section*{Results}
\label{sec:results}
In this section I report the results from executing Scenarios~1-4 from Section~\ref{sec:scenarios} using TensorFlow~\cite{TensorFlow:Lib}, version 1.12, on a 2017 MacBook Pro with 3.1 GHz Intel Core i7 and 16 GB RAM {\em without} Graphics Processing Unit (GPU) acceleration. The test code is written in Python 3 and executes under a Docker image~\cite{TensorFlow:Install} configured with 10 GB of RAM and 4 GB swap. 

\paragraph {Scenario 1.} In this test, the BowTow neural network is tested with encoding~\ref{eq:multihot}. The results in Table~\ref{tab:Scenario1} show high accuracy and low binary cross-entropy estimates. 
\begin{table}[!ht]
\begin{adjustwidth}{0in}{0in} 
\centering
\caption{{\bf Scenario 1 results.} Data from experiments with training a model on KID until it attains some validation accuracy grater than 88 \%. Note that each time the data is loaded for training, it is shuffled randomly, hence the small variation in computational results.}
\begin{tabular}{|c|c|}
\hline
\multicolumn{2}{|c|}{\bf Training and validating on KID} \\ \hline
validation accuracy (\%)& validation binary cross-entropy\\ \hline
88.08 & 0.2955\\ \hline
88.18 & 0.2887\\ \hline
88.21 & 0.2945\\ \hline
\end{tabular}
\label{tab:Scenario1}
\end{adjustwidth}
\end{table}

To assess the relative computational efficiency of the BowTie neural network, I compared it to the convolutional neural network in~\cite{Keras-Conv-Examples} with a 10 000 word dictionary. The network reached accuracy of 88.92 \% at Epoch 4 with binary cross entropy of 0.2682. However, it took 91 seconds/Epoch, which matched the numbers reported by the authors for the CPU-only computational platform. In addition, after Epoch 4, the binary cross-entropy started to increase steadily while the accuracy started to decline. For example, the binary cross-entropy reached 0.4325 at Epoch 10 with validation accuracy of 87.76 \% and 0.5536 and 87.40 \% correspondingly at Epoch 15. 

\begin{wrapfigure}[14]{l}{70mm}
\includegraphics[width=80mm]{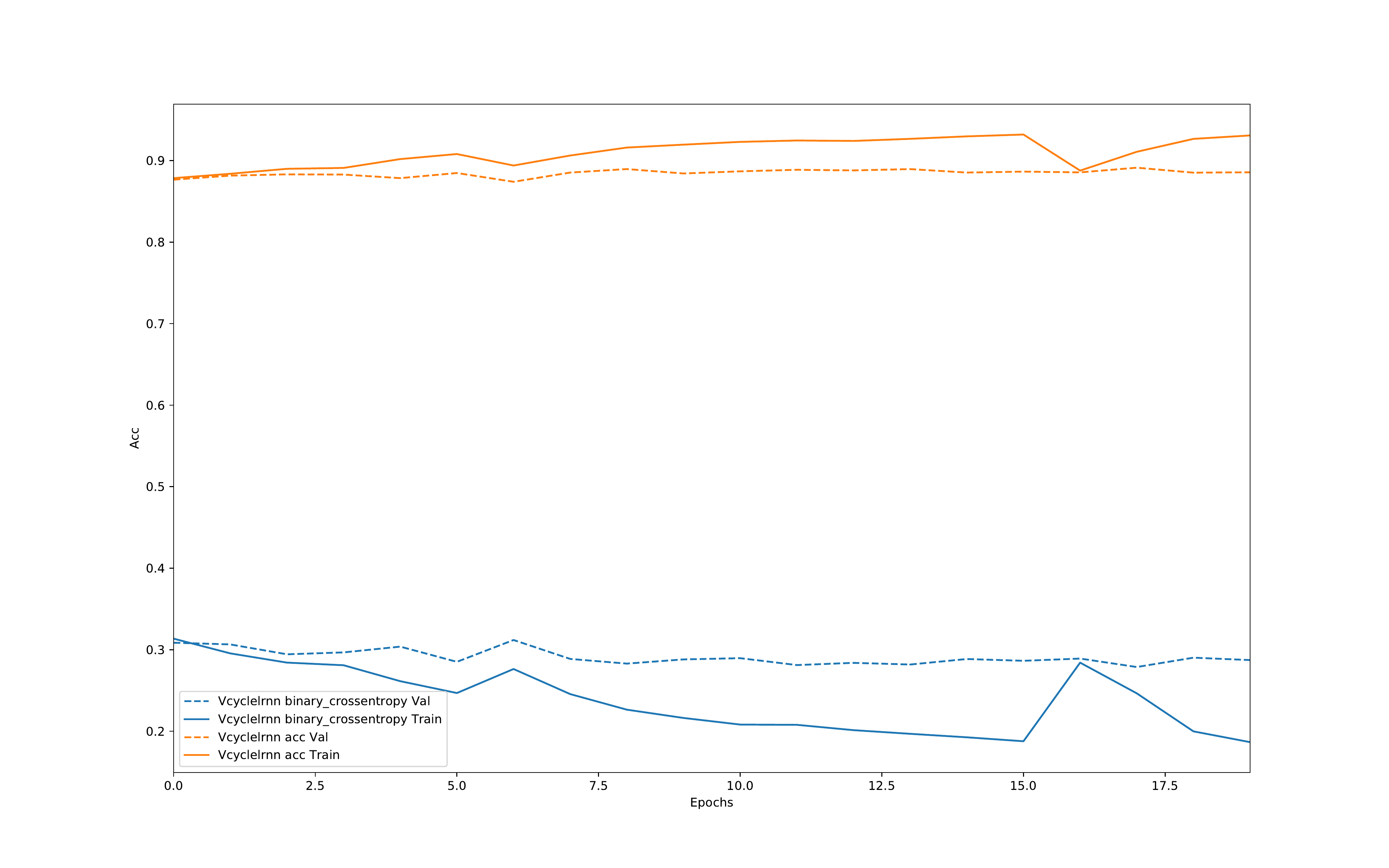}
\captionsetup{labelformat=empty} 
\caption{} 
\label{fig:BowTieResults} 
\end{wrapfigure}

\marginpar{
\vspace{.1cm} 
\color{Gray} 
\textbf{Figure \ref{fig:BowTieResults}. BowTie accuracy and cross-entropy results.} 
The neural network keeps the cross-entropy estimate low over the course of 20 epochs.
}
In comparison, BowTie took only 3 seconds/Epoch for the same dictionary size and attained accuracy of 88.20 \% with binary cross-entropy of 0.2898. The binary cross-entropy stays stable below 0.38 for a number of Epochs. In other words, BowTie is 30-times faster, attains comparable accuracy and maintains stable binary-cross entropy.   

\paragraph{Scenario 2} SLMRD is more challenging than KID for reasons that are visible in the test results for Scenarios~3 and 4, hence the slightly lower validation accuracy attained by BowTie using the simple encoding~\ref{eq:multihot} - it easily meets or exceeds the threshold accuracy of 87.95 \% but could not surpass the 88 \% limit in several experiments. 

\begin{table}[!ht]
\begin{adjustwidth}{0in}{0in} 
\centering
\caption{{\bf Scenario 2 results.} Data from experiments with training a model on KID until it attains some validation accuracy was equal to or grater than 87.95 \%. Note that each time the data is loaded for training, it is shuffled randomly, hence the small variation in computational results.}
\begin{tabular}{|c|c|}
\hline
\multicolumn{2}{|c|}{\bf Training and validating on SLMRD} \\ \hline
validation accuracy (\%)& validation binary cross-entropy\\ \hline
87.98 & 0.2959\\ \hline
87.95 & 0.2996\\ \hline
87.96 & 0.3001\\ \hline
\end{tabular}
\label{tab:Scenario2}
\end{adjustwidth}
\end{table}

\begin{wrapfigure}[14]{l}{70mm}
\includegraphics[width=80mm]{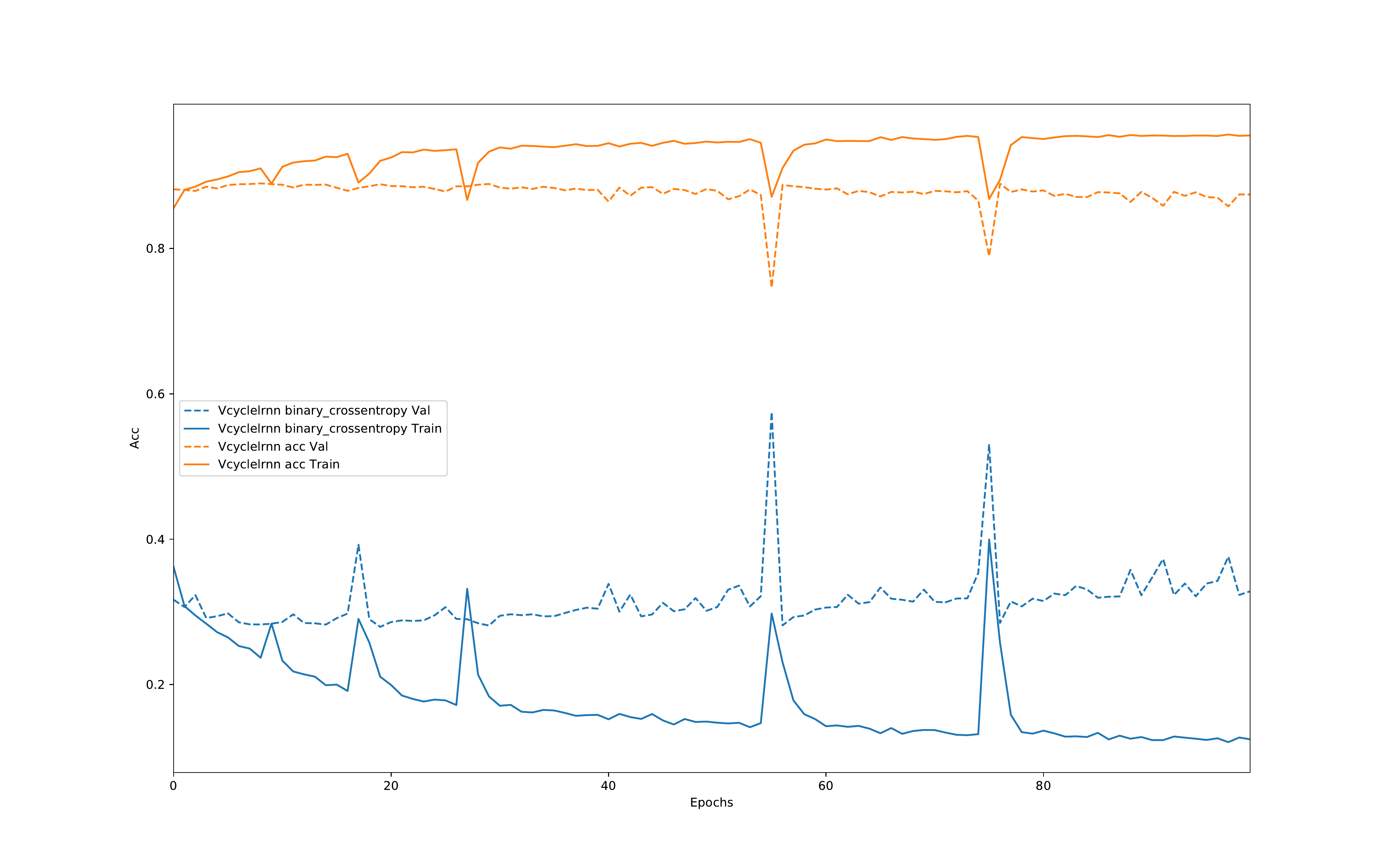}
\captionsetup{labelformat=empty} 
\caption{} 
\label{fig:BowTie100Epochs} 
\end{wrapfigure}

\paragraph{Scenarios 3 and 4.}I combine the reporting for Scenarios~3 and 4 because once the model is trained under Scenario~3 it is then transferred to compute predictions on KID. To perform the transfer testing on KID one needs to reconcile the difference in $\vert\Pi_{SLMRD}\vert$ and $\vert\Pi_{KID}\vert$. As I noted in Section~\ref{sec:data}, $\vert\Pi_{SLMRD}\vert=89\, 527$ and $\vert\Pi_{KID}\vert=88 587$. Moreover, $\Pi_{KID}\not\subset\Pi_{SLMRD}$. Let $\Pi_{\Delta}=\Pi_{KID}\setminus (\Pi_{KID}\cap\Pi_{SLMRD})$. It turns out that
\begin{center}
$\Pi_{\Delta}=\left\{
\begin{array}{l}0’s, 1990s, 5, 18th, 80s, 90s, 2006, 2008, \\85, 86, 0, 5, 10, tri, 25, 4, 40’s, 70’s, 1975, \\1981, 1984, 1995, 2007, dah, walmington, \\19, 40s, 12, 1938, 1998, 2, 1940's, 3, 000, 15, 50.\end{array}\right\}.$ 
\end{center}  
Clearly, $\vert\Pi_{\Delta}\vert$ is small and of the words in $\Pi_{\Delta}$, only 'walmington' looks like a plausible English word but it is the name of a fictional town in a British TV series from the 1970’s. As such it has no computable polarity index and is semantically negligible. Based on this, I dropped all these words during the mapping of $\pi\in\Pi_{\Delta}$ into the corresponding $\pi\prime\in\Pi_{SLRMD}$. Note that this $\pi\rightarrow\pi\prime$ mapping enables the encoding of the semantics of the texts in KID according to \ref{eq:polarity-weighted-multihot}.   
\begin{paragraph}
{Some simple but revealing statistics about the data.} With encoding \ref{eq:polarity-weighted-multihot}, the matrix $T_{SLMRD}$ for the training set in SLMRD shows cumulative polarity in the range [-50.072 837, 58.753 546] and the cumulative polarity of the elements of the matrix $T_{SLMRD}$ for the test set is in the range [-48.960 107, 63.346 164]. This suggests that SLMRD is pretty neutral and an excellent dataset for training models. In contrast, the elements of the matrix $T_{KID}$ are in the range [-49.500 000, 197.842 862].      
\end{paragraph}
\begin{table}[!ht]
\begin{adjustwidth}{-0.2in}{0in} 
\centering
\caption{{\bf Scenarios 3 and 4 results.} Data from experiments with training a model on SLMRD until it attains some validation accuracy grater than 89 \% and then using that same model to predict the category for each labeled review in KID. Note that the transfer accuracy is computed over the entire set of 50 000 reviews in KID.}
\begin{tabular}{|c|c|c|}
\hline
\multicolumn{2}{|c|}{\bf Training on SLMRD} & {\bf Predicting on KID}\\ \hline
validation accuracy (\%)& validation binary cross-entropy& Prediction transfer accuracy (\%)\\ \hline
89.02 & 0.2791 & 91.56 \\ \hline
89.12 & 0.2815 & 91.63 \\ \hline
89.17 & 0.2772 & 91.76 \\ \hline
\end{tabular}
\label{tab:Sec3-4}
\end{adjustwidth}
\end{table}
\newpage
\marginpar{
\vspace{.1cm} 
\color{Gray} 
\textbf{Figure \ref{fig:BowTie100Epochs}. BowTie accuracy and cross-entropy results.} 
The neural network keeps the cross-entropy estimate in check over the course of 100 epochs.
}
Table~\ref{tab:Sec3-4} contains the results from executing Scenarios~3 and 4.
The validation and transfer accuracy results in Table~\ref{tab:Sec3-4} are better than those shown in Tables~\ref{tab:Scenario1} and \ref{tab:Scenario2}. This shows the value word polarity brings in capturing the semantics of the text. The transfer accuracy over KID is also higher than the results obtained by using the convolutional neural network~\cite{Keras-Conv-Examples} on KID. 

The results in Table\ref{tab:Sec3-4} are higher than the results reported for sentiment prediction of movie reviews in \cite{Conneau-at-al-2017} but also in agreement with the reported experience by these authors of consistently improved accuracy from supervised learning on a large representative corpus before transferring the model for prediction to a corpus of interest.  

Note also that the validation accuracy results for the BowTie neural network on SLMRD are substantially higher than the results for the same dataset in~\cite{maas-EtAl:2011:ACL-HLT2011}. The observation in \cite{maas-EtAl:2011:ACL-HLT2011} that even small percentage improvements result in a significant number of correctly classified reviews applies to the data in Table~\ref{tab:Sec3-4}: that is, there are between 172 and 210 more correctly classified reviews for BowTie. In addition, BowTie is computationally stable and retains low cross-entropy losses over a large number of epochs, see Figures~\ref{fig:BowTieResults} and \ref{fig:BowTie100Epochs}, which is a desirable property.     

The speed of computation improves on platforms with GPU acceleration. For example, experiments on a system {\em with} eight Tesla V100-SXM2 GPUs yield speedups of nearly a factor of two in training but the acceleration plateaued if more than two GPUs were used. The computational speedup was better during prediction computations with a trained model: the time needed to calculate prediction for the entire set of 50 000 reviews reduced from 44 secs on one GPU to 31 secs on two GPUs, to 28 secs on four GPUs and to 26 secs on eight GPUs.  


\section*{Discussion and next steps}

The experimental results from sentiment prediction presented above show the great potential deep learning has to enable automation in areas previously considered impossible. At the same time, we are witnessing troubling trends of deterioration in cybersecurity that have permeated the business and home environments: people often cannot access the tools they need to work or lose the data that is important to them~\cite{Verizon-BIR}. Traditionally, governments and industry groups have approached this problem by establishing security testing and validation programs whose purpose is to identify and eliminate security flaws in IT products before adopting them for use. 

\begin{wrapfigure}[13]{l}{75mm}
\includegraphics[width=75mm]{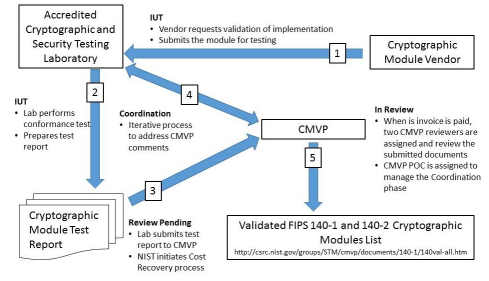}
  \begin{picture}(0.0, 0.0)
    \put(160.0,120){\includegraphics[height=0.3in]{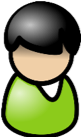}}
    \put(55.0,128){\includegraphics[height=0.3in]{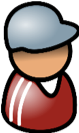}}
    \put(115.0,85){\includegraphics[height=0.3in]{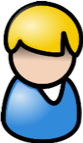}}
    \put(130.0,125.0){\includegraphics[height=0.25in]{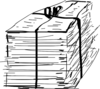}}
    \put(100.0,90.0){\includegraphics[height=.2in]{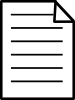}}
    \put(15.0,52.0){\includegraphics[height=0.25in]{document-pile-th}}
  \end{picture}
\captionsetup{labelformat=empty} 
\caption{} 
\label{fig:CMVP} 
\end{wrapfigure}

One area of specific concern in cybersecurity is cryptography.     
Society recognizes cryptography's fundamental role in protecting sensitive information from unauthorized disclosure or modification. The cybersecurity recommendations in \cite{Verizon-BIR} list relying on cryptography as a means of data protection as one of the top recommendations to the business community for the past several years.  However, the validation programs to this day remain heavily based on human activities involving reading and assessing human-written documents in the form of technical essays - see Fig.~\ref{fig:CMVP} for an illustration of the structure of the Cryptographic Module Validation Program (CMVP)~\cite{2017:NIST:CMVP} established in 1995 to validate cryptographic modules against the security requirements in Federal Information Processing Standard (FIPS) Publication \mbox{140-2}~\cite{2001:NIST:FIPS-140-2}.

This validation model worked well for the level of the technology available at the time when the programs were created more than two decades ago.  As technology has advanced, however, this model no longer satisfies current day industry and government operational needs in the context of increased number and intensity of cybersecurity breaches~\cite{Verizon-BIR}.

\marginpar{
\vspace{.7cm} 
\color{Gray} 
\textbf{Figure \ref{fig:CMVP}. CMVP structure and processes} 
Processes rely entirely on human actors and human-readable artifacts (English essays)
} 
There are several factors for this. First, current cybersecurity recommendations~\cite{Verizon-BIR} require that every organization relying on technology today patch promptly, including applying patches to cryptographic modules. Technology products are very complex and the cost of testing them fully to guarantee trouble-free use is prohibitively high. As a result, products contain vulnerabilities that hackers and technology producers are competing to discover first: the companies to fix, the hackers to exploit.  Patching products changes the game for hackers and slows down their progress. Thus, patching promptly is a way of staying ahead of security breaches.  However, patching changes also the environment in which a cryptographic module runs and may also change the module itself, thus invalidating the previously validated configuration. Users who depend on validated cryptography face a dilemma when frequent updates and patches are important for staying ahead of the attackers, but the existing validation process does not permit rapid implementation of these updates while maintaining a validated status because of the slow human-based activities illustrated in Fig.~\ref{fig:CMVP}. 

The second factor hindering the effectiveness of the traditional validation model shown in Fig.~\ref{fig:CMVP} is the demand for speed in the context of the cognitive abilities of the human brain. Rapid changes in technology and related steep learning curves are stretching the resources at the testing laboratories. When evaluation package submissions finally reach the validation queue, inconsistent and possibly incomplete evidence presentation further strains the ability for a finite number of reviewers to provide timely turnaround. Recent scientific research points out that humans are limited in their ability to process quickly and objectively large amounts of complex data~\cite{kahneman2011thinking}. 
Two systems drive the way humans think: the fast, intuitive, and emotional  System 1; the slower, more deliberative, and more logical System 2. Systems 1 and 2 constantly interact but System 1 is always in the driver's seat. This leads to faults and biases, of fast thinking, and reveals the pervasive influence of intuitive impressions on human thoughts and behavior. These are the primary reasons why we cannot trust human intuitions when dealing with highly complex data.

Going back to the results on sentiment analysis with deep learning from above and in spite of that success people may always question the ability of machines to replace humans in solving such cognitive and analytical tasks. They will always ask why the accuracy is not one hundred percent? Or, notwithstanding the available scientific evidence~\cite{kahneman2011thinking},  say that if a human was reviewing the text she would have never made a mistake. 
        
Changing public opinion may be a slow process. Besides, it seems that cybersecurity and machine learning/artificial intelligence (AI) will always be joined at the hip because the more we rely on machines to solve ever more complex tasks, the higher the risk those machines may be attacked to subvert their operation. Can AI fight back though? The successful results presented in this paper suggest that computer-based validation of cryptographic test evidence may be the only viable alternative that would allow objective and accurate assessment of large volumes of data at the speed required by the cybersecurity reality~\cite{Verizon-BIR}. This paper demonstrates that deep learning neural networks are capable of tackling the core tasks in security validations and thereby automate existing programs~\cite{2018:NIST:ACVT}. If this effort is indeed successful one may reason that by helping to improve the process of validation and thereby increase cybersecurity, albeit indirectly, AI will in fact be defending itself from cybersecurity threats. Over time, this may lead to societal acceptance of AI into sensitive domains such as the validation of critical components for the IT infrastructure.

\paragraph{Next steps.}
The importance of incorporating word polarity into the model is illustrated clearly by the results presented in this paper. However, the type of language used in the validation test reports tends to be different than the colloquial English used in movie reviews. The technical jargon in test reports uses many common words whose meaning changes in this context. Moreover, the assessments for each test requirement in \cite{2011:NIST:FIPS-140-2-DTR} are written in a way very different from movie reviews. Here the author provides arguments that justify her conclusion about compliance. The challenge is to distinguish weak/faulty arguments from solid ones. Therefore, a new type of polarity needs to be developed. Related to that is the task of assembling a representative corpus of labeled validation test report data for training and validation.     


\section*{Acknowledgments}
I thank the NIST Information Technology Laboratory (ITL) and especially Elham Tabassi for the research funding and support under ITL Grant \#7735282-000.

\nolinenumbers

\bibliography{bowtie}

\begin{thebibliography}{10}

\bibitem{Stanford:LMRD}
{Andrew Maas}.
\newblock Large movie review dataset.
\newblock \url{http://ai.stanford.edu/~amaas/data/sentiment/}, 2011.

\bibitem{Bramble-1993}
J.~H. Bramble.
\newblock {\em Multigrid Methods}.
\newblock Wiley, 1993.
\newblock Pitman Research Notes in Mathematics Vol. 294.

\bibitem{Brants-at-al-2007}
T.~Brants, A.~C. Popat, P.~Xu, F.~J. Och, and J.~Dean.
\newblock Large language models in machine translation.
\newblock In {\em Proceedings of the 2007 Joint Conference on Empirical Methods
  in Natural Language Processing and Computational Natural Language Learning
  (Prague)}, pages 858--867, June 2007.

\bibitem{Conneau-at-al-2017}
A.~Conneau, D.~Kiela, H.~Schwenk, L.~Barraul, and A.~Bordes.
\newblock Supervised learning of universal sentence representations from
  natural language inference data.
\newblock In {\em Proceedings of the 2017 Conference on Emprical Methods in
  Natural Language Processing (Copenhagen, Denmark, September 7-11))},
  Association of Computational Linguistics, pages 670--680, 2017, See also
  update at \url{https://arxiv.org/abs/1705.02364v5}.

\bibitem{Goodfellow-et-al-2016}
I.~Goodfellow, Y.~Bengio, and A.~Courville.
\newblock {\em Deep Learning}.
\newblock MIT Press, 2016.
\newblock \url{http://www.deeplearningbook.org}.

\bibitem{TensorFlow:Lib}
{Google LLC}.
\newblock Tensorflow: An open source machine learning framework for everyone.
\newblock \url{https://www.tensorflow.org/}, 2019.

\bibitem{TensorFlow:Install}
{Google LLC}.
\newblock Tensorflow: Docker.
\newblock \url{https://www.tensorflow.org/install/docker}, 2019.

\bibitem{Hayou-at-al-2019}
S.~Hayou, A.~Doucet, and J.~Rousseau.
\newblock On the impact of the activation function on deep neural networks
  training.
\newblock \url{https://arxiv.org/abs/1902.06853}, 2019.

\bibitem{kahneman2011thinking}
D.~Kahneman.
\newblock {\em Thinking, fast and slow}.
\newblock Farrar, Straus and Giroux, New York, 2011.

\bibitem{Keras:IMDB}
{Keras Documentation}.
\newblock Imdb movie reviews sentiment classification.
\newblock \url{https://keras.io/datasets/}, 2018.

\bibitem{Keras-Conv-Examples}
{Keras Team}.
\newblock Demonstration of the use of convolution1d for text classification.
\newblock
  \url{https://github.com/keras-team/keras/blob/master/examples/imdb_cnn.py},
  2019.

\bibitem{maas-EtAl:2011:ACL-HLT2011}
A.~L. Maas, R.~E. Daly, P.~T. Pham, D.~Huang, A.~Y. Ng, and C.~Potts.
\newblock Learning word vectors for sentiment analysis.
\newblock In {\em Proceedings of the 49th Annual Meeting of the Association for
  Computational Linguistics: Human Language Technologies}, pages 142--150,
  Portland, Oregon, USA, June 2011. Association for Computational Linguistics.

\bibitem{Mikolov-at-al-2013}
T.~Mikolov, I.~Sutskever, K.~Chen, G.~Corrado, and J.~Dean.
\newblock Distributed representations of words and phrases and their
  compositionality.
\newblock In {\em Advances in neural information processing systems}, pages
  3111--3119, 2013, see also \url{https://arxiv.org/abs/1310.4546}.

\bibitem{Nacson-at-al-2018}
M.~S. Nacson, J.~Lee, S.~Gunasekar, P.~H.~P. Savarese, N.~Srebro, and
  D.~Soudry.
\newblock Convergence of gradient descent on separable data.
\newblock \url{https://arxiv.org/abs/1803.01905v2}, 2018.

\bibitem{2001:NIST:FIPS-140-2}
{NIST}.
\newblock Security requirements for cryptographic modules, federal infomation
  processing standard ({FIPS}) 140-2.
\newblock \url{https://doi.org/10.6028/NIST.FIPS.140-2}, 2001.

\bibitem{2011:NIST:FIPS-140-2-DTR}
{NIST}.
\newblock Derived test requirements for fips pub 140-2, security requirements
  for cryptographic modules.
\newblock
  \url{https://csrc.nist.gov/CSRC/media/Projects/Cryptographic-Module-Validation-Program/documents/fips140-2/FIPS1402DTR.pdf},
  2011.

\bibitem{2018:NIST:ACVT}
{NIST}.
\newblock Automated cryptographic validation testing.
\newblock \url{https://csrc.nist.gov/projects/acvt/}, 2018.

\bibitem{2017:NIST:CMVP}
{NIST}.
\newblock {Cryptographic Module Validation Program}.
\newblock
  \url{https://csrc.nist.gov/projects/cryptographic-module-validation-program},
  2018.

\bibitem{Potts-2011}
C.~Potts.
\newblock On the negativity of negation.
\newblock In {\em Proceedings of Semantics and Linguistic Theory}, volume~20,
  pages 636--659, 2011.

\bibitem{Verizon-BIR}
Verizon.
\newblock 2018 data breach investigations report.
\newblock \url{https://enterprise.verizon.com/resources/reports/dbir/}, 2018.

\end{thebibliography}

\bibliographystyle{abbrv}

\end{document}